\documentclass[
    ,final            % use final for the camera ready runs
  ]
  {aipproc}

\layoutstyle{6x9}
\usepackage{epsfig}
\begin{document}

%\title{Higher Twist, $\xi_w$ Scaling,  and 
%Effective $LO~PDFs$  for Lepton 
%Scattering in the Few GeV Region}
\title{Modeling Neutrino and Electron Scattering Cross
Sections in the Few GeV 
Region with  
Effective $LO~PDFs$}

\author{A Bodek}{
  address={Department of Physics and Astronomy,  University of Rochester,
Rochester, New York 14618,  USA}
}

\author{ U K Yang}{
  address={Enrico Fermi Institute, University of Chicago,Chicago,
 Illinois 60637, USA}
}

%\author{<author3>}{
%  address={<common address for author2 and author3>}
%  ,altaddress={<author1 address>} % additional visiting address
%}

\begin{abstract}
%This template file shows how to use the \texttt{aipproc} class to
 We use new scaling variables  $x_w$ and $\xi_w$,  and add low $Q^2$
modifications to GRV94 and GRV98 leading order
          parton distribution functions such that
          they can be used to model electron, muon and
          neutrino  inelastic scattering cross sections (and
	  also photoproduction) at both
	  very low and high energies ( $Invited$ $talk$ $given$ $by$ $Arie$ $Bodek$
	  $at$ $the$ $X$ $Mexican$ $School$ $of$
$Particles$ $and$ $Fields,$ $Playa$ $del$ $Carmen,$ $Mexico,$ $2002$) 
	  
\end{abstract}

\maketitle

%%%%%%%%%%%%%%%%%%%%%%%%%%%%%%%%%%%%%%%%%%%%
%% MAINMATTER
%%%%%%%%%%%%%%%%%%%%%%%%%%%%%%%%%%%%%%%%%%%%

%\section{Origin of Higher Twist Terms}

  The quark distributions in the proton
and neutron are parametrized as
Parton Distribution Functions (PDFs) obtained from
global fits to various sets of data at very high energies.
These fits are done within
the theory of Quantum Chromodynamics (QCD) in either leading order
(LO) or next to leading order (NLO). The most important data
come from deep-inelastic
e/$\mu$ scattering experiments on hydrogen and deuterium,
and $\nu_\mu$ and $\overline\nu_\mu$ experiments on nuclear targets.
In previous publications~\cite{highx,nnlo,yangthesis}
we have compared the predictions of the
NLO MRSR2 PDFs to deep-inelastic e/$\mu$ scattering data~\cite{slac}
on hydrogen
and deuterium from SLAC, BCDMS and
NMC. In order to get agreement with the lower
energy SLAC data  for
 $F_2$ and $R$ down to $Q^{2}$=1 GeV$^2$, and
at the highest values of $x$ ($x = 0.9$), we found
that the following modifications  to the NLO MRSR2 PDFs must be
included.
\begin{enumerate}
       \item The relative normalizations between the various
             data sets and the BCDMS systematic error shift must
              be included~\cite{highx,nnlo}.
        \item Deuteron binding corrections need to be applied and
%         discussed in ref.~\cite{highx}.
%        \item 
	the ratio of $d/u$ at high $x$ must be increased as
          discussed in ref.~\cite{highx}.
\item Kinematic higher-twist originating from target mass effects~\cite{gp}
are very large and
must be included.
\item Dynamical higher-twist corrections are smaller but also need
          to be included~\cite{highx,nnlo}.
\item In addition, our  analysis including QCD Next to NLO
(NNLO) terms shows~\cite{nnlo} that most of the dynamical 
higher-twist corrections
 needed to fit the data within a NLO QCD analysis  originate from
the $missing$ $NNLO$ $higher$ $order$ $terms$.
\end{enumerate} 
Our analysis shows that the NLO MRSR2 PDFs with target
mass and NNLO higher order terms describe electron and muon
scattering $F_2$ and $R$ data with a very small contribution from higher twists.
Studies by other authors~\cite{Blum} also show that in NNLO
analyses the dynamic higher twist
corrections are very small.
If (for $Q^2>$ 1 GeV$^2$) most of the higher-twist terms needed to
obtain agreement with the low energy data actually originate from target mass
effects and missing NNLO terms (i.e. not from interactions with
spectator quarks) then these
terms should be the same in  $\nu_\mu$  and e/$\mu$  scattering.
Therefore, low energy  $\nu_\mu$  data
should be described by the PDFs which are fit to high energy
data and are modified to include target mass and
higher-twist corrections that fit 
low energy e/$\mu$ scattering data.  However,  for $Q^2<$ 1 GeV$^2$
additional non-perturbative effects from spectator quarks must also be 
included~\cite{first}.

\section{Previous Results with GRV94 PDFs and $x_w$}

In a previous communication~\cite{first}
we used a modified scaling variable $x_w$ and fit for modifications
 to the GRV94
 %~\cite{grv94}
 leading order PDFs such that the PDFs
describe both high energy low energy e/$\mu$ data. In order to describe
 low energy data down to the photoproduction
limit ($Q^{2} = 0$), and account for both target mass and higher twist effects,
the following modifications of the GRV94 LO PDFs are need:

\begin{figure}[t]
%\centerline{\psfig{figure=nufact_fig2.ps,width=5.5in,height=5.4in}}
\centerline{\psfig{figure=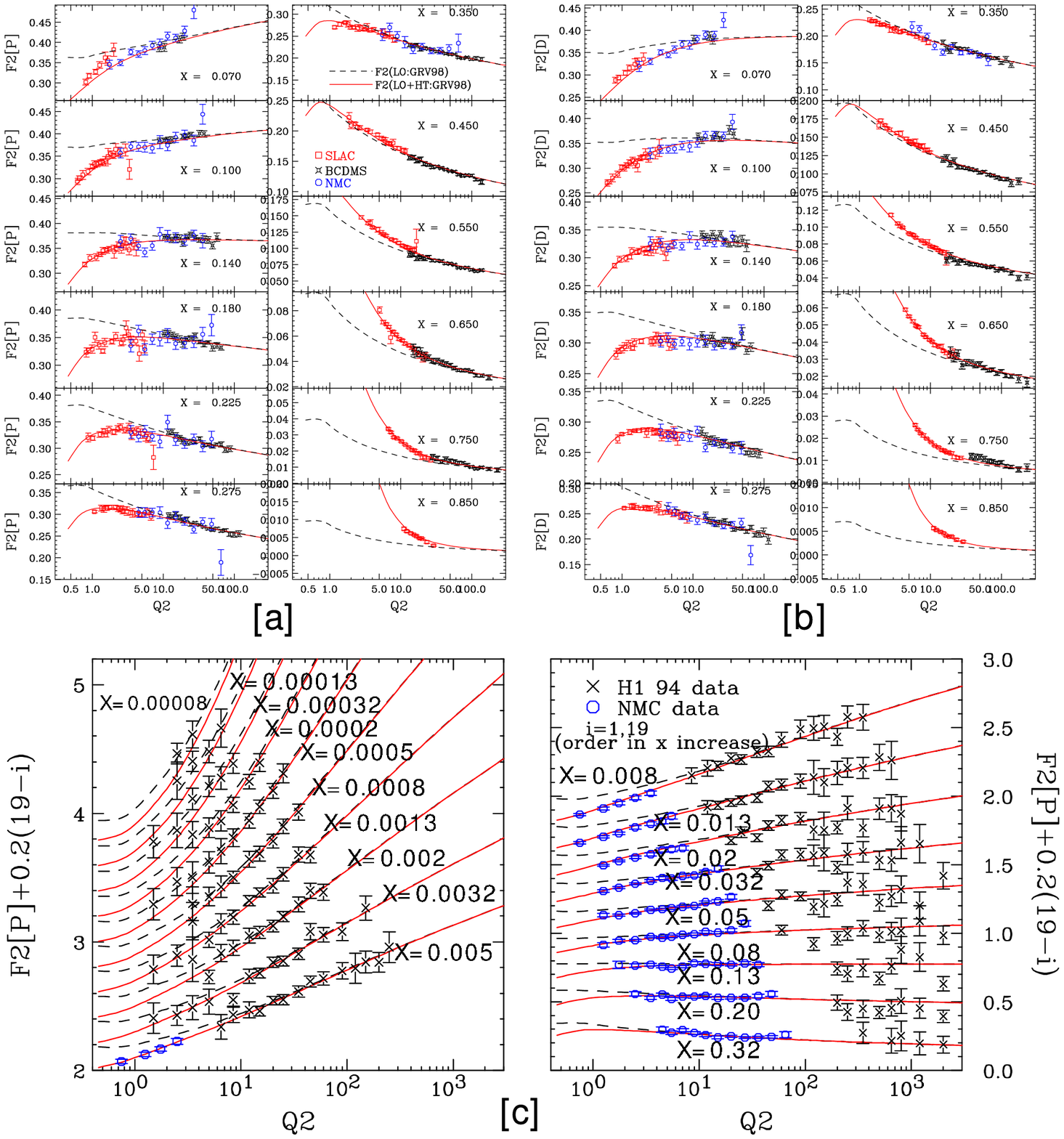,width=5.5in,height=5.4in}}
\caption{Electron and muon $F_2$ data (SLAC, BCDMS, NMC, H1 94)
used in our  GRV98 $\xi_w$ fit compared to the predictions of the unmodified GRV98
PDFs (LO, dashed line) and the modified GRV98 PDFs 
fits (LO+HT, solid line); [a] for $F_2$ proton, [b] for $F_2$ deuteron,
and [c] for the H1 and NMC proton data at low $x$.}
\label{fig:f2fit}
\end{figure}

\begin{enumerate}
\item  We increased the $d/u$ ratio at high $x$ as described in our previous
analysis~\cite{highx}.
\item  Instead of
the scaling variable $x$ we used the
scaling variable $x_w = (Q^2+B)/(2M\nu+A)$ (or =$x(Q^2 +B)/(Q^2+Ax)$).
This modification was used in early fits to SLAC data~\cite{bodek}.
The parameter A provides for an approximate way to include $both$ target
mass and higher twist effects at high $x$,
and the parameter B allows the fit to be
used all the way down to the photoproduction limit ($Q^{2}$=0).
\item  In addition as was done in earlier non-QCD based
fits~\cite{DL} to low energy data, we multiplied all PDFs
by a factor $K$=$Q^{2}$ / ($Q^{2}$ +C). This was done in order for
the fits to describe low $Q^2$
 data in the photoproduction limit, where 
$F_{2}$ is related to the
photoproduction cross section according to
\begin{eqnarray}
     \sigma(\gamma p) = {4\pi^{2}\alpha_{\rm EM}\over {Q^{2}}}
          F_{2}
           = {0.112 mb~GeV^{2}\over {Q^{2}}}   F_{2} \nonumber
\label{eq:photo}
\nonumber
\end{eqnarray}
\item Finally,  we froze
 the evolution of the GRV94 PDFs at a
value of $Q^{2}=0.24$ (for $Q^{2}<0.24$),
because GRV94 PDFs are only valid down to $Q^{2}=0.23$ GeV$^2$.
\end{enumerate}

In our analyses, the measured structure functions
were corrected for the BCDMS systematic error shift and for
the relative normalizations between the SLAC, BCDMS
and NMC data~\cite{highx,nnlo}.
The deuterium data were corrected
for nuclear binding effects~\cite{highx,nnlo}. 
A simultaneous fit  to both proton and deuteron SLAC, NMC and BCDMS data
(with $x>0.07$ only)
yields A=1.735, B=0.624 and C=0.188 (GeV$^2$) with GRV94 LO PDFs
($\chi^{2}=$ 1351/958 DOF). 
Note that for $x_w$ the parameter
A  accounts for $both$ target mass and higher twist effects.
\begin{figure}[t]
%\centerline{\psfig{figure=figure2Mex1.eps,width=6.0in,height=3.6in}}
\centerline{\psfig{figure=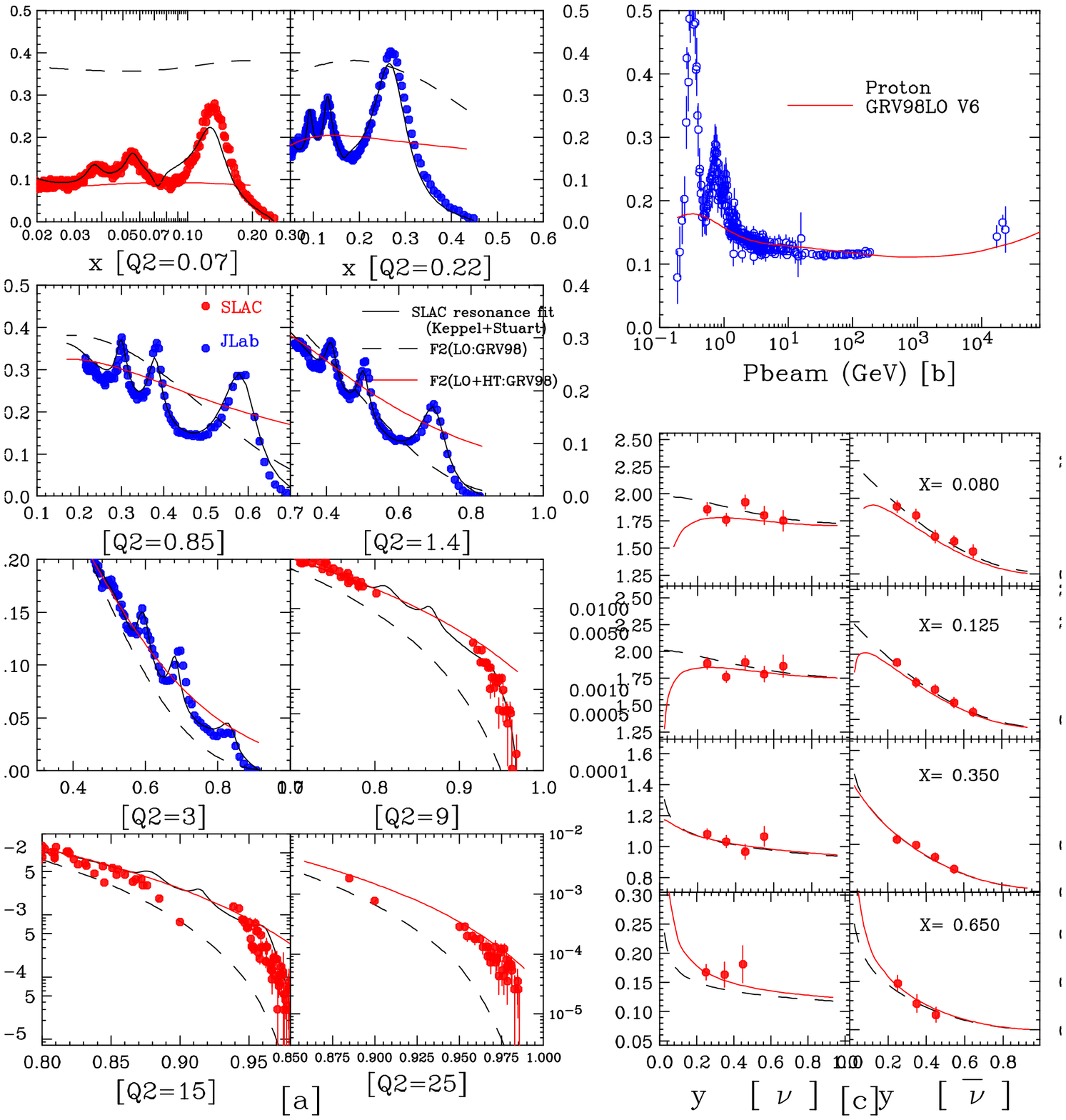,width=6.0in,height=3.6in}}
\caption{ Comparisons to proton and
iron data not included in our GRV98 $\xi_w$  fit. 
(a) Comparison of SLAC and JLab 
(electron) $F_{2p}$ data
in the resonance region (or fits to these data)
and the predictions of the GRV98 PDFs with (LO+HT, solid)
and without (LO, dashed) our modifications.
(b)  Comparison of photoproduction
       data on protons  to predictions using
our modified GRV98 PDFs. 
(c)  Comparison of representative CCFR $\nu_\mu$ and $\overline\nu_\mu$
charged-current differential cross sections~\cite{yangthesis,rccfr}
on iron at 55 GeV and the predictions of the GRV98 PDFs with (LO+HT,
solid) and without (LO, dashed)
      our modifications. 
}
\label{fig:predict}
\end{figure}
\section{New Analysis with $\xi_w$, $G_D$ and GRV98 PDFs }
In this
publication we update our previous studies,~\cite{second}
which were done with a new improved scaling variable  $\xi_w$, and fit for
modifications to the more modern GRV98  LO PDFs such that the PDFs
describe both high energy and low energy electron/muon data.
We now also include  NMC and H1 94 data at lower $x$.
Here  we freeze
the evolution of the GRV98 PDFs at a
value of $Q^{2}=0.8$ (for $Q^{2}<0.8$),
because GRV98 PDFs are only valid down to $Q^{2}=0.8$ GeV$^2$.
In addition, we use different photoproduction limit
multiplicative factors for valence and sea. Our proposed
new scaling variable is based on the following derivation.
Using energy momentum conservation, it can be shown that
the factional momentum $\xi$ = $(p_z + p_0)/(P_z + P_0)$
carried by a quark of 4-mometum $p$ 
in a proton target of mass M and 4-momentum P is given
by  $\xi$ = $xQ^{'2}/[0.5Q^{2}(1+[1+(2Mx)^{2}/Q^2]^{1/2})]$, where

$2Q^{'2} =[Q^2+M_f{^2}-M_i{^2}]+[(Q^2+M_f{^2}-M_i{^2})^2 
+4Q^2(M_i{^2}+P_{T}^{2})]^{1/2}.$

Here $M_i$ is the initial quark mass with average initial
transverse momentum $P_T$ and $M_f$ is the mass of the
quark in the final state.  The
above expression for $\xi$ was previously derived~\cite{gp}
for the case of $P_T=0$. Assuming $M_i=0$ we use instead:  

$\xi_w = x(Q^2+B + M_f{^2})/(0.5Q^{2}(1+[1+(2Mx)^{2}/Q^2]^{1/2})+Ax)$

Here $M_f$=0, except for charm-production processes in neutrino scattering
for which $M_f$=1.5 GeV.
%The $\xi_w$ scaling variable  includes the exact form
%for the target
%mass effects~\cite{gp} in the dominator. 
 For $\xi_w$ the parameter A
is expected to be much smaller than for  $x_w$ since now it only  
accounts for the  higher order (dynamic higher twist) QCD terms 
in the form of an $enhanced$ target mass term (the effects of the proton
target mass are already taken into account using the exact form
in the denominator of $\xi_w$ ).
The parameter
B accounts for the initial state quark transverse
momentum and final state quark  $effective$ $\Delta M_f{^2}$
 (originating from multi-gluon emission by quarks).

%For the valence quarks, we improve on the
%low $Q^2$ multiplicative factor $K$ as follows.
Using closure considerations~\cite{close} ($e.g.$the Gottfried
sum rule) it can be shown
that, at low $Q^2$, 
the scaling prediction for the $valence$
quark part of $F_2$ should be multiplied by
the factor $K$=[1-$G_D^{2}$($Q^{2}$)][1+M($Q^{2}$)] where
$G_D$ = 1/(1+$Q^{2}$/0.71)$^{2}$ is the  proton elastic form factor,
and M($Q^{2})$ is related to the magnetic elastic form factors
of the proton and neutron.
At low $Q^2$, [1-$G_D^{2}$($Q^{2}$)]
is approximately $Q^{2}$/($Q^{2}$ +C) with  $C=0.71/4=0.178$ (versus
 our fit value C=0.18 with GRV94).  In order to
 satisfy the Adler Sum rule~\cite{adler} we add the function M($Q^{2}$)
to account for  terms from the magnetic 
and axial elastic form factors of the nucleon).
Therefore, we try a more general form  
$K_{valence}$=[1-$G_D^{2}$($Q^{2}$)][$Q^{2}$+$C_{2v}$]/[$Q^{2}$ +$C_{1v}$],
and   $K_{sea}$=$Q^{2}$/($Q^{2}$+$C_{sea}$).  Using this 
 form with the GRV98 PDFs (and now also including
 the very low $x$ NMC and H1 94 data in the fit)  we find
 $A$=0.419, $B$=0.223, and  $C_{1v}$=0.544, $C_{2v}$=0.431, and $C_{sea}$=0.380
(all in GeV$^2$, $\chi^{2}=$ 1235/1200 DOF).
As expected, A and B are now smaller with
respect to our previous fits with GRV94 and $x_w$.
 With these modifications, the GRV98 PDFs must also be multiplied
by $N$=1.011 to $normalize$ to the SLAC $F_{2p}$ data.
The fit (Figure  \ref{fig:f2fit}) yields the
following  normalizations  relative to
the SLAC $F_{2p}$ data ($SLAC_D$=0.986, $BCDMS_P$=0.964, $BCDMS_D$=0.984,
$NMC_P$=1.00, $NMC_D$=0.993, $H1_P$=0.977, and BCDMS systematic error shift of 
1.7).(Note, since the GRV98 PDFs do not include the charm sea,
for $Q^{2}>0.8$ GeV$^2$
we add charm production using the photon-gluon fusion
model in order to fit the very high $\nu$ HERA data. This is not
needed for any of the low energy comparisons but is only needed
to describe the highest $\nu$ HERA electro and photoproduction data). 

\begin{figure}[t]
%\centerline{\psfig{figure=
%f2d_grv98-model_duality_xsi.ps,width=5.5in,height=3.2in}}
\centerline{\psfig{figure=
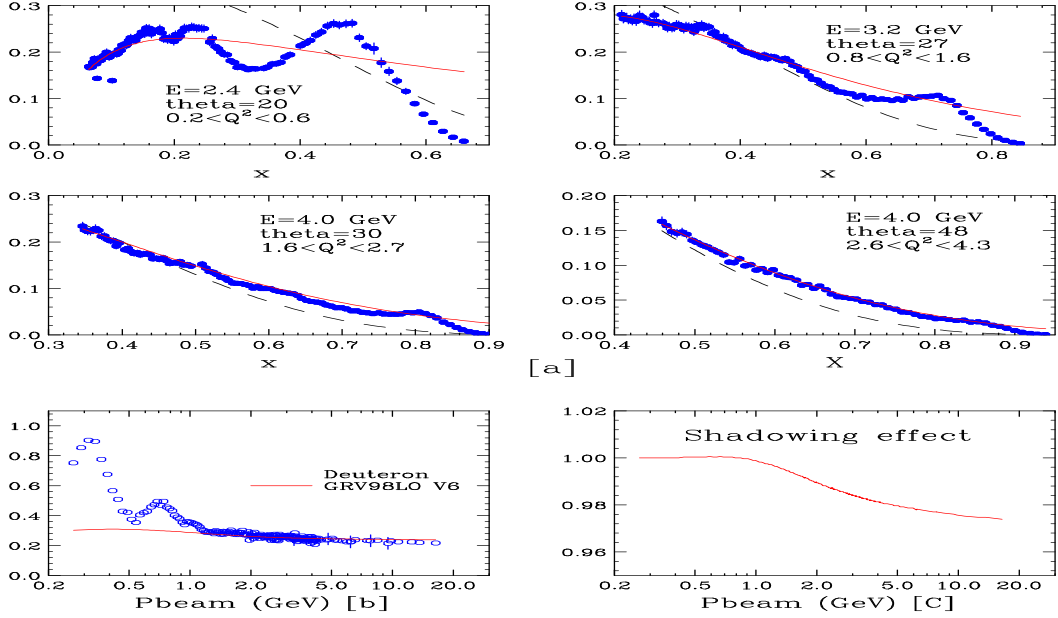,width=5.5in,height=3.2in}}
\caption{ Comparisons to data on deutrerium
which were not included in our GRV98 $\xi_w$ fit. 
(a) Comparison of SLAC and JLab 
(electron) $F_{2d}$ data
in the resonance region 
and the predictions of the GRV98 PDFs with (LO+HT, solid)
and without (LO, dashed) our modifications.
(b)  Comparison of photoproduction
       data on deuterium  to predictions using
our modified GRV98 PDFs (including shadowing corrections). 
(c) The  shadowing corrections that were applied to the PDFs for
predicting the photoproduction cross section on deuterium.}
\label{fig:predictD}
\end{figure}

Comparisons of $predictions$
using these modified GRV98 PDFs to other data which were $not$  
$ included$ 
in the fit is shown in Figures  \ref{fig:predict} and \ref{fig:predictD}.
From duality~\cite{bloom} considerations,
with the $\xi_w$ scaling variable, the modified
GRV98  PDFs should
also provide a reasonable
description of the average value of $F_2$
 in the resonance region. Figures \ref{fig:predict}(a) and \ref{fig:predictD}(a) show
a comparison between resonance data (from SLAC and Jefferson
Lab, or parametrizations of these data~\cite{jlab}) on protons and
deuterons versus
the predictions with the standard  GRV98 PDFs (LO)
and with our modified GRV98 PDFs (LO+HT). The 
modified GRVB98 PDFs are in good agreement with SLAC and JLab
resonance data down to $Q^{2}=0.07$ (although resonance
data were not included in our fits). 
There is also very
good agreement
of the  $predictions$ of our modified GRV98 
in the $Q^2=0$ limit with
 photoproduction  data on protons and deuterons  as shown in
Figure  \ref{fig:predict}(b)  and \ref{fig:predictD}(b). 
In predicting the photoproduction cross sections on deuterium,
we have applied shadowing corrections~\cite{shadow}
as shown in Figure  \ref{fig:predictD}(c). 
We also compare the
$predictions$ with  our modified GRV98 PDFs (LO+HT) to a few
representative high energy CCFR
$\nu_\mu$ and $\overline\nu_\mu$ charged-current
differential cross sections~\cite{yangthesis,rccfr} on iron
(neutrino data were not included in our fit).
In this comparison we use the
PDFs to obtain $F_{2}$ and $xF_{3}$ and correct for nuclear effects
in iron~\cite{first}.
The structure function $2xF_{1}$
is  obtained by using the $R_{world}$ fit from reference~\cite{slac}.
There is very good agreement of our $predictions$
 with these neutrino data on iron. 

In order to have a full description of all charged current
$\nu_\mu$ and $\overline\nu_\mu$
processes,  the contribution from quasielastic
scattering~\cite{qe} must be added
separately at $x=1$.   The best prescription is to
use our model in the region
 above the first resonance (above $W$=1.35 GeV) and
add the contributions  from
quasielastic and first resonance~\cite{rs} ($W$=1.23 GeV) separately.
This  is because the  $W=M$ and  $W$=1.23 GeV regions are dominated by
one and two isospin states, and the amplitudes for neutrino
versus electron scattering are related via Clebsch-Gordon rules~\cite{rs}
instead of quark charges (also the V and A
couplings are not equal at low $W$ and $Q^2$).
In the region of higher mass resonances 
      (e.g. $W$=1.7 GeV) there is a significant
contribution from the deep-inelastic continuum which is not
well modeled by the existing fits~\cite{rs} to
neutrino resonance data (and using our modified 
PDFs should be better).
For nuclear targets, nuclear corrections~\cite{first} must also be applied.
Recent results from Jlab indicate that the Fe/D ratio in the
resonance region 
is the same as the Fe/D ratio from DIS data for the same value of
$\xi$ (or $\xi_w$).
The effects of terms proportional to the muon mass and $F_4$ and 
$F_5$ structure functions in neutrino scattering are small and
are discussed in Ref.~\cite{qe,kr}.
In the  future, we plan to investigate the effects of including
the initial
state quark $P_T$ in $\xi_w$,
and institute further improvements such as allowing for
different higher twist parameters for u, d, s, c, b  
quarks in the sea, and the small difference
(expected in the Adler sum rule) in the $K$ factors
for axial and vector terms in neutrino scattering.
In addition,
we can multiply the PDFs by a modulating
function~\cite{bodek,close} A(W,$Q^{2}$)  to improve  modeling  in
the resonance region (for hydrogen) by including 
(instead of $predicting$)
the resonance  data~\cite{jlab} in the
fit.
We can also include resonance 
data on deuterium~\cite{jlab} and heavier nuclear targets
in the fit, and low energy neutrino data.
Note that because
of the effects of experimental resolution and Fermi motion
~\cite{fermi}
(for nuclear targets), a description of the average cross
section in the resonance region is sufficient for most neutrino
experiments.

\section{Appendix}

In leading order QCD (e.g.
GRV98 LO PDFs), $F_2$ for the scattering
of electrons and muons on proton (or neutron) targets is
given by the sum of quark
and anti-quark distributions (each weighted the
square of the quark charges):
\begin{eqnarray}
F_2(x) &=& \Sigma_i e_i^2 \left [xq_i(x)+x\overline{q}_i(x) \right]\\
2xF_1(x) &=& F_2(x) (1+4Mx^2/Q^2) / (1+R_{world}).
\end{eqnarray}
%In the extraction of original
%GRV98 LO PDFs, no separate longitudinal contribution was
%included. The quark distributions
%were directly fit to $F_2$ data. A full
%modeling of electron and muon cross section requires
%also a description of $2xF_1$.
%We use a non-zero longitudinal $R$ in reconstructing $2xF_1$
%by using a fit of $R$ to measured data.
%Thus, $2xF_1$ is given by
%\begin{eqnarray}
%2xF_1(x) &=& F_2(x) (1+4Mx^2/Q^2) / (1+R_{world}).
%\end{eqnarray}
%
Here, $R_{world}$ is parameterized~\cite{slac} by:
\begin{eqnarray}
R_{world}(x,Q^2) & = & \frac{0.0635}{log(Q^2/0.04)} \theta(x,Q^2)
%\nonumber \\
	          + \frac{0.5747}{Q^2}-\frac{0.3534}{Q^4+0.09},
\end{eqnarray}
where $\theta = 1. + \frac{12 Q^2}{Q^2+1.0}
              \times \frac{0.125^2}{0.125^2 + x^2}$.
%UK

The $R_{world}$ function provides a good
description of the world's data in the $Q^2>0.5$ and $x>0.05$ region.
Note that the $R_{world}$ function breaks down
below $Q^2=0.3$. Therefore, we freeze the function at $Q^2=0.35$ and introduce
the following function for $R$ in the $Q^2<0.35$ region.
The new function provides a smooth transition
from $Q^2=0.35$ down to $Q^2=0$ by forcing $R$ to approach zero at $Q^2=0$
as expected in the photoproduction limit (while
keeping a $1/Q^2$ behavior at large $Q^2$ and matching to $R_{world}$
at  $Q^2=0.35$).
\begin{eqnarray}
R(x,Q^2) & = & 3.207 \times \frac {Q^2}{Q^4+1}
%\nonumber \\
           \times R_{world}(x,Q^2=0.35).
\label{eq:rmod}
\end{eqnarray}
%In neutrino scattering the value of $R$ is required to approach zero
%at $Q^2=0$ only for the vector
%part of the interaction.
%However, the overall contribution
%from $R$ is expected to be small in this region. Therefore,
%  it is reasonable to use Eq.~\ref{eq:rmod} for $R$ in
%  both the electron/muon and neutrino scattering cases.

In the comparison with CCFR charged-current differential cross section
on iron, a nuclear correction for iron targets is applied.
We use the following parameterized function, $f(x)$
(fit to experimental electron and muon
scattering data for the ratio of iron to deuterium cross sections),
to convert deuterium structure functions to (isoscalar) iron
structure functions~\cite{first};
\begin{eqnarray}
f(x) & = & 1.096 -0.364x - 0.278e^{-21.94x}
%\nonumber \\
     +2.772x^{14.417}
\end{eqnarray}

For the ratio of deuterium cross sections to cross
sections
on free nucleons we use the following function
obtained from a fit to SLAC data on the nuclear
dependence of electron scattering cross sections~\cite{yangthesis}.
\begin{eqnarray}
f(x) & = & (0.985 \pm 0.0013)\times (1+0.422x-2.745x^2 \nonumber \\
        &   & +7.570x^3  -10.335x^4+5.422x^5).
\label{eq:nucl-d}
\end{eqnarray}
This correction  is only valid
in the $0.05<x<0.75$ region.  In neutrino scattering,
we use the same nuclear correction factor for $F_{2}$, $xF_{3}$
and $2xF_{1}$.

The $d/u$ correction for the GRV98 LO PDFs is obtained
from the NMC data for $F_2^D/F_2^P$.
Here, Eq.~\ref{eq:nucl-d} is used to remove nuclear binding effects
in the NMC deuterium $F_2$ data. The correction term, $\delta (d/u)$
is obtained
by keeping the total valence and sea quarks the same.
\begin{eqnarray}
\delta (d/u)(x)& = & -0.00817 + 0.0506x + 0.0798x^2,
\end{eqnarray}
where the corrected $d/u$ ratio is $(d/u)'=(d/u)+\delta (d/u)$.
Thus, the modified $u$ and $d$ valence distributions are given by
\begin{eqnarray}
u_v' = \frac{u_v}{1+\delta (d/u) \frac{u_v}{u_v+d_v}} \\
d_v' = \frac{d_v+u_v \delta (d/u)}{1+\delta (d/u) \frac{u_v}{u_v+d_v}}.
\end{eqnarray}
The same formalism is applied to the modified $u$ and $d$ sea distributions.
Accidently, the modified $u$ and $d$ sea distributions (based on NMC data)
agree with the NUSEA data in the range of $x$ between 0.1 and 0.4.
Thus, we find that any futher correction on sea quarks is not necessary.

%% If the bibliography is produced without BibTeX comment out the
% AB THE FIRST CHOICE WAS USED IN THE TEMPLATE I Ccommented both temp

%commented out line below
%\bibliography{sample}

%FOR NOW USE Old Referecnes procedure
%\section*{References}

\end{document}